\documentclass[prd,preprint,superscriptaddress]{revtex4}
\usepackage{revsymb}
\usepackage{amsmath}
\usepackage{graphicx}
\newcommand{\be}{\begin{equation}}
\newcommand{\ee}{\end{equation}}
\newcommand{\ben}{\begin{eqnarray}}
\newcommand{\een}{\end{eqnarray}}
\newcommand{\n}{\label}
\newcommand{\no}{\noindent}
\newcommand{\ov}{\overline}
\begin{document}
\title{An exact model of conformal quintessence}
\author{Luis P. Chimento\footnote{Electronic mail address: chimento@df.uba.ar}}
\affiliation{Departamento de F\'{\i}sica, Universidad de Buenos Aires,
1428 Buenos Aires, Argentina}
\author{Alejandro S. Jakubi\footnote{Electronic mail address jakubi@df.uba.ar}}
\affiliation{Departamento de F\'{\i}sica, Universidad de Buenos Aires,
1428 Buenos Aires, Argentina}
\author{Diego Pav\'on\footnote{Electronic mail address: diego.pavon@uab.es}}
\affiliation{Departamento de F\'{\i}sica, Facultad de Ciencias,
Universidad Aut\'onoma de Barcelona, 08193, Bellaterra (Barcelona), Spain}
\begin{abstract}
A non--minimally coupled quintessence model is investigated and the conditions
for a stationary solution to the coincidence problem are obtained. For a
conformally coupled scalar field and dissipative matter, a general solution
possessing late acceleration is found. It fits rather well the high redshift
supernovae data and gives a good prediction of the age of the Universe.
Likewise, the cold dark matter component dominates the cosmological
perturbations at late times albeit they decrease with expansion.
\end{abstract}


\maketitle

\section{Introduction}
After some years of research, the accelerated expansion of the Universe
appears to have gained further ground \cite{tegmark} but the nature of dark
energy -the substratum behind this acceleration- remains as elusive as ever
\cite{iap,mars}.  While one may expect that one or other of the dark energy
candidates (a very small cosmological constant \cite{alessandro}, quintessence
\cite{caldwell}, Chaplygin gas \cite{kam}, tachyon field \cite{tach},
interacting quintessence \cite{st,intq}, non-minimally coupled quintessence
\cite{nonmc,smash}, etc) will finally emerge as the successful model, at the
time being none of them is in position to claim such status.

In introducing dark energy as a novel component most of these candidates
encounter the so--called ``coincidence problem,'' namely, ``why are the energy
densities of both components (dark energy and dark matter) of the precisely
same order today?" As shown by the authors, this problem has a dynamical
solution provided that the dark matter component is assumed to be dissipative
\cite{enlarged} or interacts with the dark energy \cite{st}.  In such a case,
it can be demonstrated that the equations governing the cosmic evolution imply
the stationary condition (Eq. (\ref{attractor}) below) and that the system is
attracted to a stationary and stable solution characterized by the constancy
of both density parameters, i.e., $\Omega_{m}$ and $\Omega_{\phi}$ tend to
constant values at late times.

The aim of this paper is to provide an exact quintessence model,
non--minimally coupled to the Ricci curvature. Non--minimal coupling naturally
arises in generalizing the Klein--Gordon equation from Minkowski space to a
curved space -for a recent review on this subject and further motivations, see
Ref. \cite{review}. We believe it is rather reasonable to explore whether a
non--minimal coupling of the scalar field acting as dark energy to the Ricci
curvature may be of help to understand the present stage of accelerated
expansion and shed some light into the nature of the dark component.  As a
first step toward weighing the contribution of the coupling to the evolution
of the Universe we shall consider the simplest case of a non--minimally
coupled quintessence with a constant potential.  As it turns out, there is a
stationary solution that ensures that for late times the ratio between both
energy densities remains a constant while the Universe asymptotically
approaches a de Sitter expansion. In this late regime, the density
perturbations of large wavelength decrease and are dark matter dominated.
Further, our results are consistent with the high redshift supernovae data and
provide a very reasonable estimate of the age of the Universe.

Section II presents our model which, aside from the non-minimally coupled
scalar field, introduces a dissipative pressure in the dark matter fluid; this
pressure turns out to be key for the solution of the coincidence problem.
Section III derives the statefinder parameters as well  as the age of the
Universe.  Section IV studies the observational constraints on our model
imposed by the high redshift supernova data. Section V investigates the long
wavelength density perturbations. Finally, section VI summarizes our findings.
As usual, a subindex zero indicates that the corresponding quantity must be
evaluated at the present time. We have chosen units so that $c= 8 \pi G = 1$.

\section{Conformally coupled scalar field}

Let us consider a Friedmann--Lema\^{\i}tre--Robertson--Walker (FLRW) spacetime
filled with two components, namely, dissipative matter and a non-minimally
coupled quintessence field. The equation of state of the first component is of
baryotropic type $p_{m} = (\gamma_{m} - 1) \rho_m$, where the baryotropic index
of matter is restricted to the range $1 \leq \gamma_m \leq 2$.  In addition to
this equilibrium pressure the matter component is assumed to have a
non--equilibrium (dissipative pressure) $\pi$ connected to entropy production.
It should be noted that barring superfluids (as Helium superfluid), this
quantity is ever--present in every matter fluid found in Nature and is
negative for expanding fluids \cite{landau}. In the case at hand, it may
either come from interactions within the dark matter, or the decay of dark
matter particles into dark particles \cite{winfried}, or from the non--linear
growth of cosmic structures \cite{dominik}, and it proves crucial to solve the
coincidence problem. Likewise, the equation of state of the quintessence
component can be written as $p_{\phi} = (\gamma_{\phi} - 1) \rho_{\phi}$ with
$\gamma_{\phi} < 1$.

The Friedmann equation and the conservation equations for the
matter fluid and quintessence read
\\
\be
3 H^{2} + 3 \frac{K}{a^{2}} =  \rho_{m} + \rho_{\phi}
\qquad (K = 1, 0, -1),
\label{feq}
\ee
\\
\be
\dot{\rho_m}+3H(\gamma_m\rho_m+\pi)=0 ,
\label{cm}
\ee
\\
\be
\n{kg}
\ddot\phi+3H\dot\phi+\xi R\phi+ \frac{\mbox{d}V(\phi)}{\mbox{d} \phi}=0,
\ee
\\
where $\xi$ denotes the non-minimal coupling constant and
the Ricci curvature scalar $R$ is related to the quintessence
scalar field by \cite{nonmc}, \cite{smash}
\\
\be
\label{t}
\left[1-\xi(1-6\xi)\phi^2\right]R=-(1-6\xi)\dot\phi^2+4V-6\xi\phi
\frac{\mbox{d}V(\phi)}{\mbox{d} \phi}+(4-3\gamma_m)\rho_m-3\pi \, .
\ee
\\
\noindent
\\
Likewise, the energy density and pressure of the quintessence field
are
\\
\be
\rho_{\phi} = \textstyle{1\over{2}} \dot{\phi}^{2} + V({\phi}) +
3 \xi \, H \phi (H \phi + 2 \dot{\phi}),
\label{valerio1}
\ee
\\
\be
p_{\phi} = \textstyle{1\over{2}} \dot{\phi}^{2} - V({\phi}) -
\xi \, \left[ 4H\phi \dot{\phi} + 2\dot{\phi}^{2} + 2 \phi \ddot{\phi}+
(2\dot H +3H^{2}) \phi^{2} \right].
\label{valerio2}
\ee
\\
Obviously, these two reduce to their minimally coupled expressions
for vanishing $\xi$.
In terms of the density parameters $\Omega_{m} \equiv \rho_{m}/(3H^{2})$,
$\Omega_{\phi} \equiv \rho_{\phi}/(3 H^{2})$ and $\Omega_{K} = - K/(aH)^{2}$,
the  set of equations (\ref{feq})--(\ref{kg}) become
\\
\be
\Omega_{m} + \Omega_{\phi} + \Omega_{K} = 1,
\label{omega}
\ee
\\
\be
\dot{\Omega}_{m} + 3H\left(\frac{2 \dot{H}}{3H^{2}} +
\gamma_{m}+ \frac{\pi}{\rho_{m}}\right) \Omega_{m} = 0,
\label{domegam}
\ee
\\
\be
\dot{\Omega}_{\phi} + 3H\left(\frac{2\dot{H}}{3H^{2}} +
\gamma_{\phi}\right) \Omega_{\phi} = 0,
\label{domegaphi}
\ee
\\
respectively.

The simplest solution to the system  of equations (\ref{domegam})-
(\ref{domegaphi}) that solves the coincidence problem is that
$\Omega_m = \Omega_{m 0}$ and $\Omega_{\phi} = \Omega_{\phi 0}$ at
late times, with $\Omega_{m 0}$ and $\Omega_{\phi 0}$ constants.
This automatically implies the stationary condition \cite{enlarged}
\\
\be
\gamma_{m}+\frac{\pi}{\rho_{m}}= \gamma_\phi=
-\frac{2\dot H}{3H^2}.
\label{attractor}
\ee
\\
It is readily seen that on the stationary solution, Eq.(\ref{attractor}),
one has $K = 0$ thereby we shall focus on spatially flat FLRW
spacetimes hence forward.

We are interested in obtaining a simplified, analytically integrable model
that still retains the essentials of non--minimally coupled approaches and
leads to an accelerated phase of expansion at late times.
This may be accomplished by choosing for the coupling constant the conformal
value $\xi = 1/6$ and a constant value for the potential, $V(\phi) = V_{0}$.
As a consequence, Eqs.  (\ref{feq}), (\ref{t}) and (\ref{kg}) take a very
simple form on the stationary solution
\\
\be
\n{00'}
3H^2=(1+r)\left[\frac{1}{2}(\dot\phi+H\phi)^2+V_0\right],
\ee
\\
\be
\n{t'}
R=4(1+r)V_{0} \, ,
\ee
\\
\be
\n{kg'}
\ddot\phi+3H\dot\phi+\frac{\omega^2}{2}\phi=0 \, ,
\ee
\\
where $ r \equiv \Omega_{m}/\Omega_{\phi}$ stands for the density
ratio, and $\omega^2 \equiv (4/3)(1+r)V_0$.  In deriving (\ref{t'}) we
have made use of the fact that the Ricci  scalar is $R =
6(\dot{H}+2H^{2})$. It is interesting to note that, notwithstanding
the potential being a constant, the gravitational interaction induces
an effective mass given by  $m_{\mathrm{eff}}^2=\omega^2/2$ in the effective
potential $V_{\mathrm{eff}}=(\omega^2\phi^2/4)+V_{0}$ of the generalized
Klein--Gordon equation. Thereby, in a loose sense, one might
associate a particle -the ``conformalon"- with the field $\phi$.

In order to integrate this system of equations it is expedient to introduce
the conformal time $\eta$ and define a new field $\psi$ as
\\
\be
\n{cv}
\eta=\int\frac{dt}{a},  \qquad     \psi=\phi \, a \, ,
\ee
\\
\no
then, Eqs. (\ref{00'})-(\ref{kg'}) become
\\
\be
\n{kg2}
3(a')^2=(1+r)\left[\frac{1}{2}(\psi')^2+a^4V_0\right],
\qquad  a''=\frac{\omega^2}{2}a^{3},
\qquad
\psi''=0 \, ,
\ee
\\
where a prime indicates derivation with respect to $\eta$.

The general solution of this system of equations is given by
\\
\be
\n{sa}
a=\sqrt{c\sinh{\omega t}},
\ee
\\
\be
\n{sc}
\phi=\frac{\psi}{a}=\frac{\sqrt{2c^2V_0}\,\eta+b}
{\sqrt{c\sinh{\omega t}}},
\ee
\\
where $c$ and $b$ are arbitrary integration constants, and the initial
singularity has been fixed at $t=0$. Combining Eq. (\ref{sa}) with the Friedmann
equation on the stationary solution, $3H^{2} = (1+r)\rho_{\phi}$, we get the
following expression for the conformal quintessence
energy density
\\
\begin{equation}  \label{rhophi}
\rho_\phi(t)=V_0 \, \mathrm{coth}^2\omega t .
\end{equation}
\\
\noindent
In the late time accelerated regime, its equation of state
\\
\begin{equation}  \label{pphi}
p_\phi(t)=\left(\frac{4}{3\cosh^2\omega t}-1\right)\rho_\phi
\end{equation}
\\
becomes that of a cosmological constant, and both density parameters
$\Omega_\phi(t)$ and $\Omega_m(t) = r \Omega_\phi(t)$
asymptotically approach constant values.

Inserting Eq. (\ref{sa}) into Eq. (\ref{cv}) it follows that
\\
\begin{equation}
\label{eta}
\eta(t)=\frac{2}{\omega\sqrt{c}}F\left[\left(1-\exp(-\omega t)\right)^{1/2},
\frac{1}{\sqrt{2}}\right],
\end{equation}
\\
\noindent
where $F$ is the elliptic integral of the first kind. The conformal time is a
growing monotonic function of $t$ that  behaves like $\sqrt{t}$ close to the
singularity  and has a finite upper bound
\\
\begin{equation}
\label{etainf}
\eta(\infty)=\frac{2K\left(1/\sqrt{2}\right)}{\omega\sqrt{c}},
\end{equation}
\\
where $K$ is the complete elliptic integral and $2K(1/\sqrt{2})\simeq 3.7$.

{}From Eq. (\ref{sc}) it is apparent that depending on the choice of the
constants several cases arise. If $b\neq 0$, then $\phi\propto
1/\sqrt{t}$ for $t\to0$ while if $b=0$, the field has an extremum
at the initial singularity.  On the other hand, at late times the
quintessence field evolves toward the minimum of its
effective potential.

\begin{figure}[c]
\includegraphics*[scale=.8]{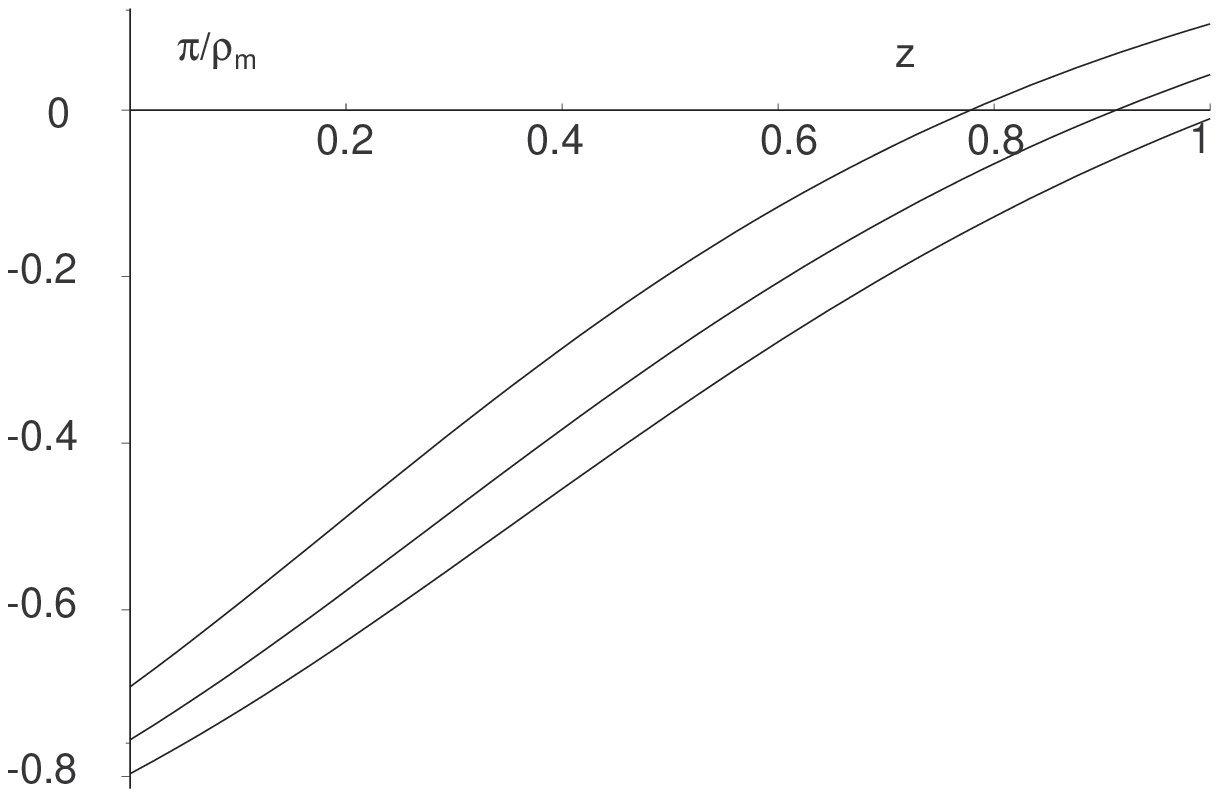}

\caption{\label{fig:piro}
Selected curves of the ratio dissipative
pressure {\em vs} energy density of matter $\pi/\rho_m$ vs the redshift $z$
between the present $z=0$ and $z=1$. From top to bottom, the curves
correspond to parameter $\sigma=0.3$, $0.224$, and $0.18$ defined
by $\sigma = (\sinh \omega t_{0})^{-2}$.
}

\end{figure}

The dissipative pressure
\\
\be
\n{pi}
\pi=\rho_m\left[-\gamma_m+\frac{4}{3\cosh^2{\omega t}}\right]
\ee
\\
follows from Eqs. (\ref{attractor}) and (\ref{sa}); and the evolution of the
ratio $\pi/\rho_{m}$, depicted in Fig. \ref{fig:piro}, shows that the relative
relevance of the dissipative pressure grows with expansion. From the last
equation, when $\gamma_m<4/3$, it is seen that  there is a critical time $t_c$
given by
\\
\begin{equation} \label{tc}
\cosh^2\omega t_c=\frac{4}{3\gamma_m}
\end{equation}
\\
that separates the epoch with $\pi>0$ from the one with  $\pi<0$. From Eqs.
(\ref{tc}) and (\ref{sa}) we see that it corresponds to the redshift \\
\begin{equation}  \label{zc}
z_c=\left[\sigma\left(\frac{4}{3\gamma_m}-1\right)\right]^{-1/4}-1 ,
\end{equation}
\\
where $\sigma\equiv(\sinh \omega t_0)^{-2}$ is a characteristic parameter of
the model with $t_{0}$ denoting the ``age'' of the Universe. As is well-known,
the second law of thermodynamics requires $\pi$ to be non--positive for
expanding fluids \cite{landau}. This means that our model is valid for
$t > t_{c}$ only ; when $t<t_{c}$ it describes a fictitious positive pressure
contribution implying that any realistic evolution of the universe would yield
a corrected value for the critical time $\tilde t_c>t_c$. Hence we would
obtain a corrected age of the Universe $\tilde t_0=t_0-t_c+\tilde t_c>t_0$.
This model may describe the growth of dissipative effects within dark matter
as a consequence of the development of density inhomogeneities \cite{dominik}.
Its characteristic growth time is given by the inverse of the conformalon
mass, i.e., by $V_{0}$ and $r$. In the late time accelerated regime, the
transport equation for the dissipative pressure becomes
$\pi\simeq -3\zeta_{b} H$. So, we obtain for the dissipative coefficient
$\zeta_{b} \simeq \gamma_{m} r[V_0/(3(1+r))]^{1/2}$
which satisfies $\zeta_{b} \geq 0$ as demanded
by the second law.

\section {Cosmological parameters}
The acceleration of the Universe is usually evaluated by the
dimensionless deceleration parameter $q = - \ddot{a}/(aH^{2})$,
where $q < 0, \, q=0, \, q>0$ describes an accelerating, a linearly
expanding (or contracting), and a decelerating  universe, respectively.
The present value $q_{0}$ does not uniquely characterize the current
accelerating phase thereby different dark energy models can lead to
the same value. Useful additional information is encoded in the
statefinder parameters $\ov{r}$ and $\ov{s}$, defined
as \cite{sahni}, \cite{alam}
\\
\be
\n{state}
\ov{r} = \frac{\dddot{a}}{aH^{3}},
\qquad  \ov{s} = \frac{\ov{r}-1}{3(q - \textstyle{1\over{2}})} \, .
\ee
\\
It is to be hoped that the pairs $(\ov{r}, q)$ and $(\ov{s}, q)$ will  provide
an accurate description of the present dynamics of the Universe and give us
some insight into the nature of dark energy. This is only natural because
$\ov{r}$ and $\ov{s}$ are directly connected to the third order term in
Taylor's expansion of the scale factor around its present
value \cite{visser}.
In the case of an expansion given by
Eq. (\ref{sa}) these parameters are found to be

\begin{equation}  \label{qrsa}
q=1-\frac{\alpha^2}{2}\,,\quad
\ov{r}=3-\frac{\alpha^2}{2}\,,\quad
\ov{s}=\frac{4-\alpha^2}{3\left(1-\alpha^2\right)} \, ,
\end{equation}

\noindent
in terms of the adimensional ratio

\begin{equation}  \label{alpha}
\alpha\equiv\frac{\omega}{H}=\frac{2}{\left[1+\sigma(1+z)^4\right]^{1/2}}\,
.
\end{equation}

\begin{figure}[c]
\includegraphics*[scale=.8]{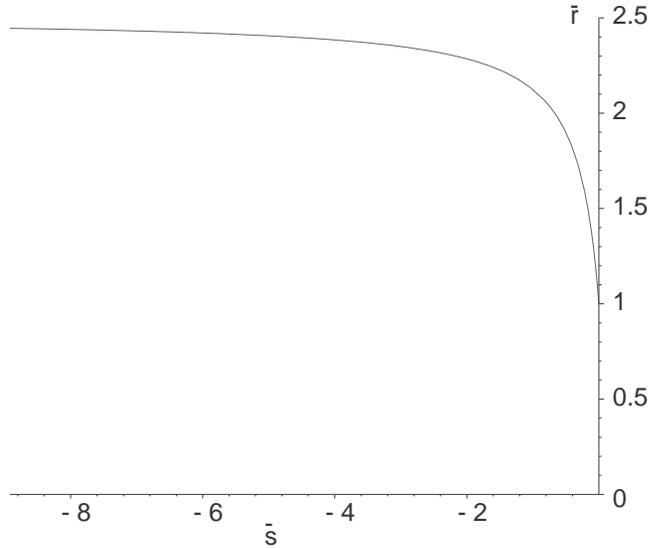}

\caption{\label{fig:rs}
Statefinder parameter $\ov{r}$ in terms of $\ov{s}$, the
other member of the pair. The evolution goes from the point
$(\ov{s}_c,\ov{r}_c)=(-\infty,5/2)$, corresponding to the critical time to the
point $(\ov{s}_\infty,\ov{r}_\infty)=(0,1)$, corresponding to the
asymptotically exponential expansion at large times. The present state of the
universe corresponds to $(\ov{s}_0,\ov{r}_0)=(-0.109\pm 0.030,1.191\pm
0.043)$.
}

\end{figure}

\begin{figure}[c]
\includegraphics*[scale=.8]{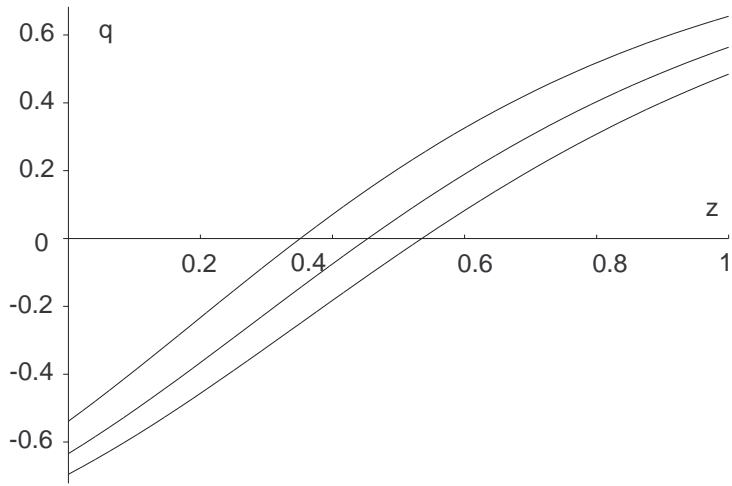}

\caption{\label{fig:q}
Selected curves of the deceleration parameter $q$ {\em vs} the redshift $z$
between the present $z=0$ and $z=1$. From top to bottom, the curves
correspond to  $\sigma=0.3$, $0.224$, and $0.18$.
}

\end{figure}

\noindent
Hence there is a single functionally independent
cosmological parameter, and the relationships
\\
\be
\n{sdr}
\ov{r} = 2+q\,, \qquad \ov{s} = \frac{2}{3} \, \frac{\ov{r} - 1}{2 \ov{r} - 5},
\ee
\\
between them hold (the dependence $\ov{r}(\ov{s})$ is depicted in
Fig. \ref{fig:rs}), so that the history of the deceleration parameter
completely describes the evolution of this universe. Since  $q \rightarrow 1$
when $t \rightarrow 0$ and $q \rightarrow - 1$ when $t \rightarrow \infty$,
this model describes a transition from a non-accelerated era to an accelerated
era in the present Universe (see Fig. \ref{fig:q}). As the accelerated phase
begins at a time $t_{ac}$, where  $\sinh(\omega t_{ac})=1$, we have that
$a_{ac}=\sqrt{c}$,

\begin{equation}  \label{tac}
t_{ac}=\frac{1}{\omega}\cosh^{-1}\sqrt{2}=
4.31\frac{\sqrt{1+\sigma}}{h} \, \mathrm{Gyr} \, ,
\end{equation}
\\
where $h$ indicates the current value of Hubble's constant in units
of $100$km/s/Mpc and the corresponding redshift is $z_{ac}=\sigma^{-1/4}-1$.
For cold dark matter ($\gamma_{m} = 1$), the ratio $\cosh(\omega
t_{ac})/\cosh(\omega t_c)=\sqrt{3/2}$ shows that at the commencement of
this phase the dissipative pressure was already negative.  On the other
hand, in virtue of Eq. (\ref{alpha}), the age of the Universe can be
expressed by
\\
\begin{equation}  \label{t0}
t_0=\frac{1}{\omega}\sinh^{-1}\frac{1}{\sqrt{\sigma}}=
4.31\frac{\sqrt{1+\sigma}}{h}\sinh^{-1}\frac{1}{\sqrt{\sigma}}\,\mathrm{Gyr},
\end{equation}
\\
implying that $\sigma$ must be lower than unity if the Universe
is to be accelerated at present. The time span since the critical
time to the present is given by
\\
\begin{equation}  \label{t0tc}
t_0-t_c=\frac{1}{\omega}\left(\sinh^{-1}\frac{1}{\sqrt{\sigma}}-
\sinh^{-1}\frac{1}{\sqrt{3}}\right).
\end{equation}

The stationary condition (\ref{attractor}) shows that
$\gamma_\phi=\gamma_m$ at the critical time. Provided that by then
the matter is cold, i.e., $\gamma_m=1$ (in the next section we will
verify the consistency of this premise) we can assume a smooth
extension of our model towards earlier times, as a cold dark matter
dominated era for the purpose of obtaining a corrected age of the
Universe. Indeed, using Eqs. (\ref{zc}), (\ref{qrsa}) and (\ref{alpha})
we find $q(z_c)=1/2$ so that the deceleration parameter of this
two-stage Universe is continuous. Hence, imposing the continuity of
the matter energy density at the critical time, we have $H_{c} =
\omega$ and $\tilde t_c=2/(3\omega)$. This yields the corrected age

\begin{equation}  \label{t0tilde}
\tilde t_0=\frac{1}{\omega}\left(\sinh^{-1}\frac{1}{\sqrt{\sigma}}-
\sinh^{-1}\frac{1}{\sqrt{3}}+\frac{2}{3}\right) .
\end{equation}

\noindent
As shown in the next section, this simple estimate produces a
rather satisfactory result.

\section{Observational constraints}
It appears that supernovae of type Ia (SNeIa) may be used as standard candles.
Properly corrected, the difference in their apparent magnitudes is related to
the cosmological parameters. Confrontation of cosmological models to recent
observations of high redshift supernovae ($z\alt 1$) have shown a good fit in
regions of the parameter space compatible with an accelerated expansion
\cite{pedagogical,Perlmutter98,Riess98,Efstat,Perlmutter99,Wang99}. We note,
however, that models like $\Lambda$CDM and QCDM usually require fine tuning to
account for the observed ratio between dark energy and clustered matter, while
our conformalon model, as well as QDDM/QIM models \cite{enlarged}, \cite{st},
simultaneously provides a late accelerated expansion and solves the
coincidence problem.

Ignoring gravitational lensing effects, the predicted magnitude for an object
at redshift $z$ in a spatially flat homogeneous and isotropic universe is
given by \cite{Peebles93}
\\
\begin{equation}
m(z) =  {\cal M} + 5\log{\cal D}_L(z),
\end{equation}
\\
\noindent
where ${\cal M}$ is its Hubble radius free absolute magnitude and
${\cal D}_L$ is the  luminosity distance in units of
the Hubble radius,
\\
\begin{equation} \label{DL}
{\cal D}_L=\left(1+z\right)\int^z_0 dz'\frac{H_0}{H\left(z'\right)}.
\end{equation}

\noindent
In virtue of Eq. (\ref{alpha}) we obtain a representation in terms of the
elliptic integral of the first kind
\\
\begin{equation}  \label{DLF}
{\cal D}_L=\frac{\sqrt{2}\left(1+z\right)\left(1+\sigma\right)^{1/2}}
{(1+i)\sigma^{1/4}}
\left[F\left(\frac{(1+i)}{\sqrt{2}}\sigma^{1/4}(1+z),i\right)-
F\left(\frac{(1+i)}{\sqrt{2}}\sigma^{1/4},i\right)\right].
\end{equation}

We have used the sample of $38$ high redshift ($0.18 \le z \le 0.83$)
supernovae of Ref. \cite{Perlmutter98}, supplemented with $16$ low redshift
($z < 0.1$) supernovae from the Cal\'an/Tololo Supernova Survey
\cite{Hamuy}. This is described as the ``primary fit'' or fit C in Ref.
\cite{Perlmutter98}, where, for each supernova, its redshift $z_i$, the
corrected magnitude $m_i$ and its dispersion $\sigma_i$ were computed.
We have determined the optimum fit of the conformalon model by minimizing a
$\chi^2$ function
\\
\begin{equation} \label{chi4}
\chi^2=\sum^N_{i=1}\frac{\left[m_i-m(z_i;\sigma,{\cal M})\right]^2}
{\sigma_i^2} \,,
\end{equation}
\\
where $N=54$ for this data set. The most likely values of these parameters are
found to be $(\sigma,{\cal M})=(0.2041,23.93)$, yielding
$\chi^2_\mathrm{min}/N_{DF}=1.104$ ($N_{DF}=52$), and a goodness--of--fit
$P(\chi^2\ge \chi^2_\mathrm{min})=0.282$. These figures show that the fit of
conformalon cosmology to this data set is even better than the fit of the
$\Lambda$CDM or QDDM/QIM models in spite of the fact that we have at our
disposal just one free parameter, namely, $\sigma$.

\begin{figure}[c]
\includegraphics*[scale=.8]{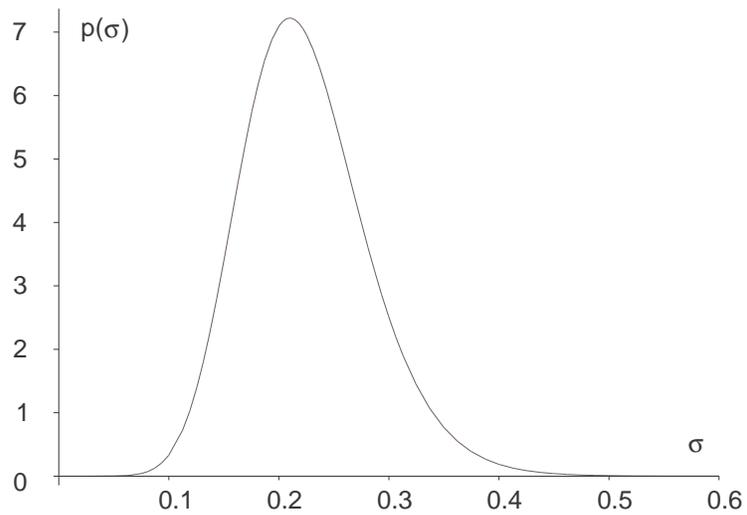}

\caption{\label{fig:ps}
The estimated probability density distribution (normalized likelihood)
for the parameter $\sigma$.
}

\end{figure}

We estimate the probability density distribution of the
parameters by evaluating the normalized likelihood
\\
\begin{equation} \label{pbetaM}
p(\sigma,{\cal M})=\frac{\exp\left(-\chi^2/2\right)}
{\int d\sigma \int d{\cal M}\exp\left(-\chi^2/2\right)} \,.
\end{equation}
\\
Then we obtain the probability density distribution for $\sigma$
marginalizing $p(\sigma,{\cal M})$ over ${\cal M}$. This probability
density distribution $p(\sigma)$ is shown in Fig. \ref{fig:ps}  and it yields
$\sigma=0.224\pm 0.054$. We next use Eq. (\ref{t0}) to obtain
$H_0t_0=0.833\pm 0.045$. Likewise, taking $h=0.7\pm 0.07$ (cf.
\cite{Krauss2003}) we get from Eq. (\ref{t0tc}) a period since the
critical time $t_0-t_c=7.46\pm 1.03 \, $Gyr, hence a corrected
age of the Universe from Eq. (\ref{t0tilde}) of $\tilde t_0= 12.7\pm
1.4 \, $Gyr.  This one standard deviation range for the corrected age of
the Universe falls within the $95\%$ confidence age range
$11.2-20 \, $Gyr derived from the age of the oldest globular clusters
and it is fully consistent with the recent estimation of $13.4 \pm
0.3 \,$ Gyr reported by the WMAP team \cite{wmap} though the latter
was reached on the basis of the standard  $\Lambda$CDM model.

By resorting to Eq. (\ref{qrsa}) we obtain $q_0=-0.809\pm 0.043$,
$\bar r_0=1.191\pm 0.043$ and $\bar s_0=-0.109\pm 0.030$. Assuming
$\gamma_m=1$ we get from Eq. (\ref{zc}) the critical redshift
$z_c=0.931\pm 0.123$. This figure is consistent with the assumption
that matter is already cold at critical time and with the  range of
redshifts of the supernovae used in the fit. On the other hand, the
accelerated expansion era begins at $z_{ac}=0.467\pm 0.093$.  This
value matches the estimation from $\Lambda$CDM and the two epoch
model of Ref. \cite{turner-riess}, where the deceleration parameter
is constant within each stage.

{}From Eq. (\ref{alpha}) we obtain the frequency parameter $\omega=0.129\pm
0.013 \, \mathrm{Gyr}^{-1}$, corresponding to a conformalon effective mass
$m_{\mathrm{eff}}=1.91\pm 0.20 \times 10^{-33} \,  \mathrm{eV}$;
and combined with the current density ratio
$r_0\simeq 0.56\pm 0.07$ \cite{flat,tytler,turner2001} yields $V_0=8.16\pm
1.69\times 10^{-3}\, \mathrm{Gyr}^{-2}$. Finally, from Eq. (\ref{pi})
we get for the current ratio
\\
\begin{equation}  \label{piro0}
\left.\frac{\pi}{\rho_m}\right|_{0}=\frac{\sigma-3}{3(1+\sigma)}=
-0.758\pm 0.051,
\end{equation}
\\
implying that nowadays the dissipative pressure plays a rather
prominent role.

\section{Cosmological perturbations}

This section considers the evolution of long-wavelength scalar perturbations
of this model. We shall follow the method employed by Perrotta and Baccigalupi
in \cite{PBM,CF} based on the formalism developed by Hwang  \cite{HW1} to
describe the evolution of perturbations in the synchronous gauge.
In this gauge the perturbed metric takes the form
\\
\begin{equation}
\label{ds2}
ds^2=a^2 [-d\eta ^2 + (\delta_{i j }+h_{i j })dx^i dx^j ]  \ ,
\end{equation}
\\
where the tensor $h_{ij}$ represents the metric perturbations and
its Fourier transform can be written as

\begin{equation}
\label{hij}
h_{i j}({\bf x},\eta )= \int {\rm d}^{3}k \, e^{i {\bf k} \cdot {\bf x}}
\left[{\bf {\hat{k}_i  \hat{k}_j}}  h( {\bf k}, \eta)
+ ( {\bf {\hat{k}_i  \hat{k}_j}}- {1 \over 3} \delta_{ i j} )
6 \zeta  ( {\bf k}, \eta) \right]\ .
\end{equation}
\\
Here $h$ denotes the trace of the tensor $h_{ij}$ and $\zeta$ represents
its traceless component -for the sake of brevity we will omit
the arguments $({\bf k},\eta )$ henceforth.

The perturbed Einstein equations read
\\
\begin{eqnarray}
\label{t00}
k^{2}\zeta -{1\over 2}{\cal H}h' &=& -{a^2 \delta \rho
\over 2}  \ ,\\
\label{ti0}
k^{2}\zeta' &=& { a^{2} (p+\rho)\theta \over 2}  \ ,\\
\label{tii}
h''+2{\cal H}h'-2k^{2}\zeta &=& -3 a^{2} \delta p \
,\\
\label{tij}
h''+6\zeta''+2{\cal H}(h'+6\zeta')-2k^{2}\zeta &=&
-3 a^{2}(p+\rho)\Sigma\ .
\end{eqnarray}
\\
where ${\cal H}=a'/a$, and the perturbed density $\delta\rho$, pressure
$\delta p$, velocity divergence $\theta$ and shear $\Sigma$ take
the form (for a detailed definition of these terms see \cite{MB,CF})
\\
\begin{equation}
\label{deltarho}
\delta \rho = {1 \over F } \left[
\delta {\rho}_{m} + {\phi' \delta\phi' \over a^2}
-{1 \over 2} F_{, \phi} \, R  \delta \phi
-3{ {\cal H} \delta F' \over a^2} -
\left(  { \rho + 3 p \over 2 } + {k^2 \over a^2} \right) \delta F +
{F'h' \over 6a^{2}}      \right] \ ,
\end{equation}
\begin{equation}
\label{deltap}
\delta p = {1 \over F } \left[ \delta p_{m} +
{\phi' \delta\phi' \over a^2}
+ {1 \over 2} F_{, \phi}\, R \delta \phi
+{\delta F''\over a^{2}} + { {\cal H} \delta F' \over a^2} +
\left(  {p- \rho  \over 2 } + {2k^2 \over 3a^2} \right) \delta F
- {F' h'  \over 9a^2}       \right]\ ,
\end{equation}
\begin{equation}
\label{theta}
(p+ \rho) \theta = {(p_{m}+ \rho_{m} ) \theta_{m}
\over F }
- {k^2 \over a^2 F } \left(
{ - \phi' \delta \phi - \delta F ' +{\cal H} \delta F } \right)\ ,
\end{equation}
\begin{equation}
\label{shear}
(p+ \rho) \Sigma=  {(p_{m}+ \rho_{m} ) \Sigma_{m}  \over  F }
+ {2k^2 \over 3a^2 F}
 \left[    \delta F + 3 { F' \over k^2} \left( \zeta' + {h'\over 6} \right)
\right]  \ ,
\end{equation}
\\
where $F=1-(\phi^{2}/6)$. There remains the perturbed Klein-Gordon equation:
\\
\begin{equation}
\label{KGpert}
\delta \phi'' + 2 {\cal H} \delta\phi' + \left( k^2
+ {1 \over 6} a^2  R
\right) \delta \phi=  {\phi' h' \over 6 } + {a^2 \over 2}
F_{, \phi} \,  \delta R \ .
\end{equation}
\\
where the perturbed Ricci scalar reads
\\
\begin{equation}
\label{deltaR}
\delta R={1\over 3a^{2}}\left(h''-3{\cal H}h'+
2k^{2}\zeta\right)\ .
\end{equation}

It has been shown in Ref. \cite{bardeen} that the density contrast
$\delta\equiv\delta\rho/\rho$ at large scales grows as $\eta^2\sim a$
during the matter dominated era previous to the critical time $\tilde
t_c$.  Here we investigate the behavior of the large scale density
perturbations in the asymptotically de Sitter era. In this regime we
have $a\simeq (c/2)^{1/2}\exp(\omega t/2)$ and
\\
\begin{equation}  \label{etaa}
\eta-\eta(\infty)\equiv \Delta\eta\simeq -\frac{2}{\omega a} \, ,
\end{equation}
\\
hence ${\cal H}\simeq -1/\Delta\eta$ and $\phi\simeq -(b\omega/2)\Delta\eta$
for $\Delta\eta\to 0^{-}$, where we have made use of Eq. (\ref{sc}) to
obtain $\phi$.
The solution of the system of equations (\ref{t00})-(\ref{deltaR}) at the
lowest order in $\Delta\eta$ and $k^2$ is readily found to be
\\
\begin{equation}  \label{peta}
h\simeq D\Delta\eta\,,\quad
\zeta\simeq -\frac{D}{6}\Delta\eta\,,\quad
\delta\rho\simeq\delta\rho_m\simeq-\delta R\simeq -\frac{\omega^2 D}{4}
\Delta\eta\,,\quad
\delta p\simeq\delta p_m\simeq\frac{\omega^2 D}{6}\Delta\eta
\end{equation}
\\
\begin{equation}  \label{dphieta}
\delta\phi\simeq A_1\Delta\eta
\end{equation}
\\
where the integration constants $D$ and  $A_1$ are functions of the wavenumber
$k$. Thus we find that matter perturbations are dominant at large times and it
holds for the perturbation of the energy density ratio $\delta
r\simeq\delta_m\equiv \delta\rho_m/\rho_m$. The density contrast decreases in
the late time regime as $\delta\propto 1/a$, so that it has a peak during the
period when viscous pressure grows.

\section{Concluding remarks}

We have presented a model of late acceleration that fits extraordinarily
well the high redshift supernovae data, yields a good prediction for
the age of the Universe and solves the coincidence problem. The only
free parameter in the supernovae fit takes a natural value, $\sigma =
0.224\pm0.054$.

The dissipative pressure $\pi$ in dark matter is a key ingredient of our
model, and it can attain comparatively large values. Such pressure may arise
from the interaction of cold dark matter with itself or the annihilation
and/or decay of this component. Different models of cold dark matter that may
show these features have been proposed recently -see, e.g., the pedagogical
short review of Ref.\cite{ostriker}. No doubt the dissipative pressure
inherent to these models may have a profound cosmological impact
\cite{physlett}.

The quintessence scalar field has a constant potential and it
is non--minimally conformally coupled to the Ricci curvature.  In our view,
the model possesses two appealing features, namely: (i) an exact and simple
solution (Eqs. (\ref{sa})-(\ref{eta})), and (ii) notwithstanding the potential
is a constant (and consequently plays the role of a cosmological constant),
the nonstandard kinetic energy term and the dissipation in the matter allows a
stationary regime where the ratio of the energy densities remains
constant. Likewise we have calculated the statefinder parameters and have
shown that on this regime they are functionally dependent so that in this
case the deceleration parameter is enough to describe these solutions.

The present model may describe the growth of dissipative effects within dark
matter with some kind of selfinteraction as a consequence of the development
of density inhomogeneities after a critical time when these inhomogeneities
become large enough. The large scale cosmological density perturbations are
seen to decrease in the asymptotic de Sitter phase, with the matter
perturbations dominating over the quintessence perturbations.

We believe that because of its simplicity, our conformal quintessence model
may serve as a starting point for more complete models that include a
nontrivial selfinteraction quintessence potential, consider the approach of
the energy density ratio towards this stationary regime, and extend towards
earlier times, when dissipative effects within dark matter become negligible.


\begin{acknowledgments}

This work has been partially supported by the Spanish Ministry of Science
and Technology under grant BFM 2000-C-03-01 and 2000-1322, and the
University of Buenos Aires under Project X223.

\end{acknowledgments}



\end{document}